\documentclass[12pt]{article}
\usepackage[dvips]{graphicx,color}
\usepackage{pdproc}
\usepackage{amsmath}
\def\b{\beta}
\def\e{\epsilon}
\def\k{\kappa}
\def\th{\theta}
\def\lam{\lambda}
\def\del{\partial}
\def\half{\frac{1}{2}}
  \makeatletter 
  \def\@cite#1{[#1]} 
  \makeatother    
\begin{document}
\renewcommand{\thefootnote}{\alph{footnote}}

\title{
Electroweak Baryogenesis in the Next-to-MSSM\footnote{
based on the collaboration 
with Koichi Funakubo~\cite{funakubo04}}
}

\author{Shuichiro Tao}

\address{ 
Department of Physics, Kyushu University\\
6-10-1 Hakozaki, Higashi-ku, Fukuoka 812-8581, Japan
\\ {\rm E-mail: tao@higgs.phys.kyushu-u.ac.jp}}

\abstract{
Next-to-MSSM is a viable model which is equipped with two features
suited for electroweak baryogensis (EWBG): first-order phase transition
not restricted by the Higgs spectrum and tree-level CP violation in 
the Higgs sector free from the nEDM constraints.
We study the model when the expectation value of the singlet scalar is of
the weak scale. Then the vacuum stability condition puts a novel upper bound
on the charged Higgs mass. We determined allowed parameter region
by requiring that the each Higgs boson is heavier than $114\mbox{GeV}$ or
its coupling to $Z$ boson is smaller than $0.1$.
}

\normalsize\baselineskip=15pt

\section{Introduction}
The MSSM has been studied by many people as a model which solves the 
gauge hierarchy problem, but there is a problem that the model contains 
a dimensional parameter $\mu$ in the superpotential. 
This problem could be avoided if we introduce a gauge singlet field, 
and consider its vacuum expected value 
as a substitution of the $\mu$-parameter.
The Next-to-MSSM (NMSSM) is the model that includes such a 
gauge singlet field $N$ in 
superpotential and soft-SUSY-breaking term 
in the following forms \cite{fayet75}:
\begin{equation}
W=-\lam NH_dH_u-\frac{1}{3}\k N^3,\quad
\mathcal{L}_{\rm soft}=-\lam A_\lam n\Phi_d\Phi_u-\frac{1}{3}\k A_\k n^3
+\text{h.c.},
\end{equation}
where we adopt $Z_3$-symmetric version of the model 
for any dimensional coupling not to exist in the superpotential. 
$H_d$ and $H_u$ are the superfields which contain the Higgs doublets
$\Phi_d$ and $\Phi_u$,  respectively.
Because the SUSY-breaking scale only characterizes this model, 
the vacuum expectation value of the singlet field $v_n$ 
is expected to be the same order as the weak scale. 
Moreover, it is also the feature of the NMSSM that 
a phase in the Higgs potential remains which cannot be removed 
by the redefinition of the fields even in the tree-level potential.

Because of the singlet Higgs field, the mass matrix of the neutral Higgs
bosons is enlarged, so that the Higgs bosons are mixed to form mass
eigenstates in a different manner from the MSSM,
\begin{equation}
\mathcal{M}^2=
\left(\begin{array}{cc}\left.\mathcal{M}^2_S\right|_{2\times2}&\\&
m^2_A\end{array}\right)
\to\left(\begin{array}{cc}\left.\mathcal{M}^2_S\right|_{3\times3}&
\\&
\left.\mathcal{M}^2_P\right|_{2\times2}\end{array}\right).
\end{equation}
For some parameter set, the coupling of the lightest Higgs boson to the 
$Z$ boson becomes very small\cite{miller03}, 
so it becomes difficult to be detected at colliders.
That is, if nature chooses this parameter set, 
$m_{h_1}$ can take a smaller value than $114$GeV.
At that time, 
it is expected that the next-lightest Higgs boson is first observed 
in the collider experiment.

Electroweak baryogenesis (EWBG) requires 
that the electroweak phase transition must be strongly first order.
This first-order phase transition could occur if 
the mass of the lightest Higgs boson $m_{h_1}$ is sufficiently light.
CP violation in the Higgs sector enhances the baryon asymmetry, since
it affects transport of (s)quarks, (s)leptons, gauginos and Higgsinos.
In the MSSM, however, the CP violation induced from the explicit CP
violation in the squark sector weakens the 
first-order phase transition\cite{funakubo03}.
On the other hand, the first-order phase transition in the NMSSM
in the case of the light Higgs will not be affected by the tree-level 
CP violation in the Higgs sector. We studied the effects of the CP violation
on the masses and couplings of the Higgs bosons.

\section{Tadpole conditions}
The tree-level Higgs potential is $V=V_F+V_D+V_{\rm soft}$. Here the
$V_F$, $V_D$ and $V_{\rm soft}$ is the following;
\begin{align}
  &V_F=|\lam n|^2(\Phi_d^\dag\Phi_d+\Phi_u^\dag\Phi_u)
  +|\e_{ij}\lam\Phi_d^i\Phi_u^j+\k n^2|^2,
  \label{eq:vfhiggs}\\
  &V_D=\frac{g_2^2+g_1^2}{8}(\Phi_d^\dag\Phi_d-\Phi_u^\dag\Phi_u)^2
  +\frac{g_2^2}{2}(\Phi_d^\dag\Phi_u)(\Phi_u^\dag\Phi_d),
  \label{eq:vdhiggs}\\
  &V_{\rm soft}=m_1^2\Phi_d^\dag\Phi_d+m_2^2\Phi_u^\dag\Phi_u+m_N^2|n|^2
  -(\e_{ij}\lam A_\lam n\Phi_d^i\Phi_u^j+\frac{1}{3}\k A_\k n^3+{\rm h.c.}).
  \label{eq:vsofthiggs}
\end{align}
Here we expand this potential around VEV which is represented by 
$v_d$, $v_u$, $v_n$ and the phases $\th$ and $\varphi$. 
The parametrization of the scalar fields are as usual,
$\Phi_d^0=(v_d+h_d+ia_d)/\sqrt{2}$, $\Phi_u^0=e^{i\th}(v_u+h_u+ia_u)/\sqrt{2}$, 
except for the singlet field,
$n = e^{i\varphi}(v_n+h_n+ia_n)/\sqrt{2}$.

The condition for the scalar potential to have a extremum at the
vacuum, is that the first derivatives with respect
to Higgs fields vanish:
\begin{equation}
 0=\frac{1}{v_i}\left<\frac{\del V}{\del h_i}\right>,\quad
 0=\frac{1}{v_i}\left<\frac{\del V}{\del a_i}\right>,\quad
(i=u,d,n).
\label{eq:tadpole-cond}
\end{equation}
These conditions are called ``tadpole conditions'' 
in the sense that 
the conditions make the tadpole diagrams vanish if we give the
vacuum expectation values $v_d$, $v_u$ and $v_n$ by hand.
Solving the second equation of (\ref{eq:tadpole-cond}), 
we get two conditions,
\begin{equation}
I_\lam=\half\mathcal{I}v_n,\quad 
I_\k=-\frac{3}{2}\mathcal{I}\frac{v_dv_u}{v_n}.
\label{eq:tadpole-p}
\end{equation}
where 
\begin{equation}
 \mathcal{I}={\rm Im}[\lam\k^*e^{i(\th-2\varphi)}],
\quad
 I_\lam=\frac{1}{\sqrt{2}}{\rm Im}[\lam A_\lam e^{i(\th+\varphi)}],
\quad
 I_\k=\frac{1}{\sqrt{2}}{\rm Im}[\k A_\k e^{i3\varphi}].
\label{eq:tadpole-im}
\end{equation}
In the Higgs sector, the phases only appear through these three combinations.
Hence our formulation so far does not depend on the convention of the phases.
Of course, this is somewhat redundant however a prospect becomes good.
Because of the two tadpole conditions on the three CP violating parameters 
$\mathcal{I}$, $I_\lam$ and $I_\k$, only one of them is the physical one.
When we introduce complex parameters, we have to manage them to 
satisfy the tadpole condition (\ref{eq:tadpole-p}).

\section{The restriction on the parameters}
\begin{figure}[htbp]
\begin{center}
\includegraphics[height=6cm]{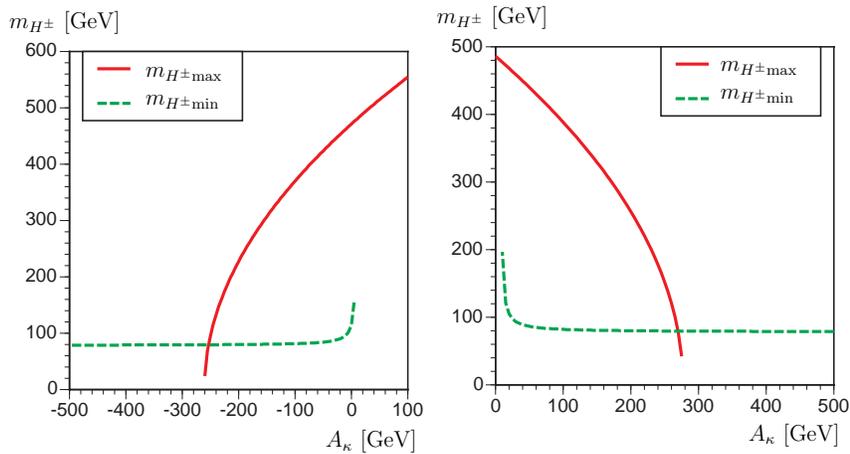}
\caption[charged Higgs bound] {
Bounds on the tree-level charged Higgs boson mass 
as a function of $A_\k$ for
$\tan\b=5$, $v_n=300{\rm GeV}$, $\lam=0.1$ and $\k=-0.3$ (left-hand plot)
and $\k=0.3$ (right-hand plot), respectively.
Solid line shows the upper mass bound of charged Higgs bosons and
dashed one shows a lower bound.}
\label{fig:charged-higgs-bound}
\end{center}
\end{figure}

Next, we explain how to restrict the parameters.
For simplicity, we assume CP-conserving case.
The necessary condition for the parameters of the NMSSM is 
that the Higgs potential must have the minimum value at the true vacuum, 
$v=246$GeV.
Although this requirement is not easy to write down unlike MSSM, 
there are two conditions which can be written down easily at tree-level; 
\begin{enumerate}
\item $V(0)>V(v_d, v_u, v_n)$, 
\item $\det\mathcal{M}_P^2>0$ 
(pseudoscalar Higgs boson must have positive mass squared),
\end{enumerate}
We adopt the physical mass of the charged Higgs boson $m_{H^\pm}$ 
as a parameter instead of the $A_\lam$ 
and now demonstrate to restrict the $m_{H^\pm}$ from above necessary conditions.
A concrete form of the conditions are as follows;
\begin{align}
 &m_{H^\pm}^2<
 \frac{2|\lam|^2v_n^2}{\sin^22\b}+\frac{2|\k|^2}{\sin^22\b}\frac{v_n^4}{v^2}
 +m_Z^2\cot^22\b+m_W^2
 +\frac{\mathcal{R}v_n^2}{\sin2\b}
 -\frac{4R_\k}{3\sin^22\b}\frac{v_n^3}{v^2},
\label{eq:charged-bound-up}
\\
&m_{H^\pm}^2>\frac{3\mathcal{R}^2v^2v_n^2}{4R_\k-3\mathcal{R}v^2\sin2\b}
+m_W^2-\half|\lam|^2v^2,
\label{eq:charged-bound-dw}
\end{align}
where
\[\mathcal{R}={\rm Re}[\lam\k^*e^{i(\th-2\varphi)}]
,\quad
R_\k=\frac{1}{\sqrt{2}}{\rm Re}[\k A_\k e^{i3\varphi}].\]
Here we plot the upper and lower bound for the 
$m_{H^\pm}$ as a function of $A_\k$
in Fig.~\ref{fig:charged-higgs-bound}.
Solid line shows the upper bound on the charged Higgs mass
which is effective when the $v_n$ is at electroweak scale and 
then suggests that the charged Higgs boson may be found in LHC,
$100<m_{H^\pm}<400$GeV. 
The dashed line shows the lower bound on the charged Higgs mass. 
The consistency of the model sets both the upper and lower bounds on $A_\k$,
which are at the weak scale when $v_n$ is also.

Actually, we search the allowed parameter sets 
using one-loop corrected Higgs potential numerically. 
Then we also demand two conditions such that the global minimum of the 
potential must be at the vacuum, and
all the scalars must have positive mass-squared,
especially the Higgs bosons which may be observed at colliders 
must be heavier than $114$GeV.
The detailed results and the analysis in the CP violating case 
are found in \cite{funakubo04}.

\section{Summary}
In the talk we try to explain the constraints for the parameters which come 
from the stability conditions of the NMSSM and also showed the example of the 
electroweak phase transition at the light-Higgs parameter set 
which is, in fact, in strongly-first order. 
The tendency of the phase transition in the NMSSM and the possibility of the 
strongly first order will be discussed in our forth-coming paper.

S.~T. and K.~Funakubo gratefully thank F.~Toyoda, A.~Kakuto and S.~Otsuki 
for valuable discussions.
(This work was supported by the Grant-in-Aid for the MEXT, 
Japan, No.~13135222.)

\bibliographystyle{plain}

\end{document}